\documentclass{article}
\usepackage{amssymb}
\usepackage{amsfonts}
\usepackage{amsmath}

\input{tcilatex}
\begin{document}

\title{Four-dimensional Brane-Chern-Simons Gravity and Cosmology}
\author{ F. G\'{o}mez$^{1,}$\thanks{%
fernando.gomez@ulagos.cl} , S. Lepe$^{2,}$\thanks{%
samuel.lepe@pucv.cl }\ and P. Salgado$^{3}$\thanks{%
patsalgado@unap.cl} \\
$^{1}$Departamento de Ciencias Exactas, Universidad de Los Lagos, \\
Avenida Fuchslocher 1305, Osorno, Chile\\
$^{2}$Instituto de F\'{\i}sica, Pontificia Universidad Cat\'{o}lica de
Valpara\'{\i}so,\\
Avenida Brasil 2950, Valpara\'{\i}so, Chile. \\
$^{3}$Instituto de Ciencias Exactas y Naturales, Facultad de Ciencias\\
Universidad Arturo Prat, Avenida Arturo Prat 2120, Iquique, Chile }
\maketitle

\begin{abstract}
From the field equations corresponding to a 4-dimensional brane embedded in
the 5-dimensional spacetime of the Einstein-Chern-Simons theory for gravity,
we find cosmological solutions that describe an accelerated expansion for a
flat universe. Apart from a quintessence-type evolution scheme, we obtain a
transient phantom evolution, which is not ruled out by the current
observational data. Additionally, a bouncing solution is shown. The
introduction of a kinetic term in the action shows a de Sitter behavior
although the energy density is not constant. \ A quintessence behavior is
also found. We conjecture on a possible geometric origin of dark energy
coming from this action.
\end{abstract}

\section{\textbf{Introduction}}

The Poincar\'{e} algebra and Poincar\'{e} group describe the symmetries of
empty Minkowski spacetime. It is known since $1970$ \cite{bacry}, that the
presence of a constant electromagnetic field in Minkowski spacetime leads to
the modification of Poincar\'{e} symmetries.

The presence of a constant classical electromagnetic field in Minkowski
spacetime modifies the Poincar\'{e} algebra into the so-called Maxwell
algebra \cite{schrader},\cite{beckers},\cite{soroka},\cite{bonanos}, \cite%
{bonanos1}, \cite{bonanos2}, \cite{gomis}. This algebra can also be obtained
from the anti-de Sitter (AdS) algebra and a particular semigroup $S$ by
means of the $S$-expansion procedure introduced in Refs. \cite{hat}, \cite%
{salg2}, \cite{salg3}, \cite{azcarr}. Using this method it is possible to
obtain more general modifications to the Poincar\'{e} algebra (see, for
example, \cite{irs}, \cite{salg4}). An interesting modification to the
Poincar\'{e} symmetries, obtained by the aforementioned expansion procedure,
is given by the so-called Lie $\mathfrak{B}$ algebra also known as
generalized Poincar\'{e} algebra, whose generators satisfy the commutation
relation shown in Eq. (7) of Ref. \cite{gomez}.

The Einstein-Chern-Simons (EChS) gravity \cite{irs} is a gauge theory whose
Lagrangian density is given by a $5$-dimensional Chern-Simons form for the $%
\mathfrak{B}$ algebra. The field content induced by the $\mathfrak{B}$
algebra includes the vielbein $e^{a}$, the spin connection $\omega ^{ab}$,
and two extra bosonic fields $h^{a}$ and $k^{ab}.$ The EChS gravity has the
interesting property that the $5$-dimensional Chern-Simons Lagrangian for
the $\mathcal{B}$ algebra, given by \cite{irs}%
\begin{eqnarray}
L_{ChS}^{(5)}[e,\omega ,h,k] &=&\alpha _{1}l^{2}\varepsilon
_{abcde}R^{ab}R^{cd}e^{e}  \label{1t} \\
&&+\alpha _{3}\varepsilon _{abcde}\left( \frac{2}{3}%
R^{ab}e^{c}e^{d}e^{e}+2l^{2}k^{ab}R^{cd}T^{\text{ }e}+l^{2}R^{ab}R^{cd}h^{e}%
\right) ,  \notag
\end{eqnarray}%
where $R^{ab}=\mathrm{d}\omega ^{ab}+\omega _{\text{ }c}^{a}\omega ^{cb}$
and $T^{a}=\mathrm{d}e^{a}+\omega _{\text{ }c}^{a}e^{c}$, leads to the
standard general relativity without cosmological constant in the limit where
the coupling constant $l$ tends to zero while keeping the Newton's constant
fixed. It should be noted that there is an absence of kinetic terms for the
fields $h^{a}$ and $k^{ab}$ in the Lagrangian $L_{ChS}^{(5)}$ (for details
see Ref. \cite{salg5}).

Recently was shown in Ref. \cite{bra1} that the $5$-dimensional EChS gravity
can be consistent with the idea of a $4$-dimensional spacetime. In this
Reference was replaced a Randall-Sundrum type metric \cite{randall} \cite%
{randall1} in the EChS gravity Lagrangian (\ref{1t}) to get (see Appendix)

\begin{eqnarray}
\tilde{S}[\tilde{e},\tilde{h}] &=&\int_{\Sigma _{4}}\tilde{\varepsilon}%
_{mnpq}\left( -\frac{1}{2}\tilde{R}^{mn}\tilde{e}^{p}\tilde{e}^{q}\right. + 
\notag \\
&&\left. +C\tilde{R}^{mn}\tilde{e}^{p}\tilde{h}^{q}-\frac{C}{4r_{c}^{2}}%
\tilde{e}^{m}\tilde{e}^{n}\tilde{e}^{p}\tilde{h}^{q}\right) ,  \label{s3}
\end{eqnarray}%
which is an gravity action with a cosmological constant for a $4$%
-dimensional brane embedded in the $5$-dimensional spacetime of the EChS
theory of gravity.\textbf{\ }$\tilde{\varepsilon}_{mnpq}$\textbf{, }$\tilde{e%
}^{m}$, $\tilde{R}^{mn}$\textbf{\ }and\textbf{\ }$\tilde{h}^{m}$\textbf{\ }%
represent, respectively, the\textbf{\ }$4$\textbf{-}dimensional versions of
the Levi-Civita symbol,\textbf{\ }the vielbein, the curvature form and a%
\textbf{\ }matter field. It is of interest to note that the field $h^{a}$, a
bosonic gauge field from the Chern-Simons gravity action, which gives rise
to a form of positive cosmological constant, appears as a consequence of
modification of the Poincar\'{e} symmetries, carried out through the
expansion procedure.

On the other hand\textbf{, }$C$\textbf{\ }and\textbf{\ }$r_{c}$\textbf{\ (}%
the "compactification radius") are constants\textbf{. }The corresponding
version in tensor language (see Appendix) is given by

\begin{equation}
\tilde{S}[\tilde{g},\tilde{h}]=\int d^{4}\tilde{x}\sqrt{-\tilde{g}}\left[ 
\tilde{R}+2C\left( \tilde{R}\tilde{h}-2\tilde{R}_{\text{ }\nu }^{\mu }\tilde{%
h}_{\text{ }\mu }^{\nu }\right) -\frac{3C}{2r_{c}^{2}}\tilde{h}\right] ,
\label{31t'}
\end{equation}%
where we can see that when $l\rightarrow 0$ then $C\rightarrow 0$ and hence (%
\ref{a}) becomes the $4$-dimensional Einstein-Hilbert action. \ 

In this paper we introduce the geometric framework obtained by gauging of
the so called\textbf{\ }$\mathfrak{B}$ algebra. Besides the vierbein $e_{\mu
}^{a}$ and the spin connection $\omega _{\mu }^{ab}$, our scheme includes
the fields\textbf{\ }$k_{\mu }^{ab}$ and $h_{\mu }^{a}$ whose dynamic is
described by the field equation obtained from the corresponding actions. The
application of the cosmological principle shows that the field $h^{a}$\ has
a similar behavior to that of a cosmological constant, which leads to the
conjecture that the equations of motion and their accelerated solutions are
compatible with the era of dark energy.

It might be of interest to note that, according to standard GR (Einstein
framework in a FLRW background), a simple way to describe dark energy (also
dark matter) is through an equation of state that relates density ($\rho $)
of a fluid and its pressure ($p$) through the equation $p=\omega \rho $,
where $\omega $ is the parameter of the equation of state. Dark energy is
characterized by $-1\leq \omega <-1/3$\textbf{, \ }$\omega =-1$ represents
the cosmological constant and $\omega <-1$ corresponds to the so-called
phantom dark energy. This means that in the context of general relativity
the parameter $\omega $ is "set by hand" and then contrasted with
observational information.

In the present work, the cosmological constant is not "set by hand" but
rather arises from the framework that we present. An example is shown where
a quintessence-type evolution as well as a phantom evolution are equally
possible. This means that a possible geometric origin of dark energy can be
conjectured in the context of the so-called Einstein-Chern-Simons gravity.

The article is organized as follows: in Section\textbf{\ }$II$\textbf{, }we
rewrite the action\textbf{\ }(\ref{31t'})\textbf{\ }by introducing a scalar
field associated to the field\textbf{\ }$\tilde{h}_{\mu \nu }$, we find the
corresponding equations of motion, and then we discuss the cosmological
consequences of this scheme. In Section\textbf{\ }$III$\textbf{, }a kinetic
term is added in the action and its effects on cosmology are studied.\
Finally, Concluding Remarks are presented in Section\textbf{\ }$IV$\textbf{. 
}An Appendix is also included where we review the derivation of the action%
\textbf{\ }(\ref{31t'}).

\section{\textbf{Cosmological consequences}}

In this Section we will study the cosmological consecuences associated with
the action\ (\ref{31t'})\textbf{. }If we consider a maximally symmetric
spacetime (for instance, the de Sitter's space), the equation 13.4.6 of Ref. 
\cite{weinberg} allows us to write the field\textbf{\ }$\tilde{h}_{\mu \nu }$%
\textbf{\ }as\textbf{\ }

\begin{equation}
\tilde{h}_{\mu \nu }=\frac{\tilde{F}(\tilde{\varphi})}{4}\tilde{g}_{\mu \nu
},  \label{66t}
\end{equation}%
where $\tilde{F}$ is an arbitrary function of an $4$-scalar field $\tilde{%
\varphi}=$ $\tilde{\varphi}(\tilde{x})$.\ This means

\begin{equation}
\tilde{R}_{\text{ }\nu }^{\mu }\tilde{h}_{\text{ }\mu }^{\nu }=\frac{\tilde{F%
}(\tilde{\varphi})}{4}\tilde{R}\text{ \ \ },\text{\ \ }\tilde{h}=\tilde{h}%
_{\mu \nu }\tilde{g}_{\text{ }}^{\mu \nu }=\tilde{F}(\tilde{\varphi}),
\end{equation}%
so that the action (\ref{31t'}) takes the form (see Appendix) 
\begin{equation}
\tilde{S}[\tilde{g},\tilde{\varphi}]=\int d^{4}\tilde{x}\sqrt{-\tilde{g}}%
\left[ \tilde{R}+C\tilde{R}\tilde{F}(\tilde{\varphi})-\frac{3C}{2r_{c}^{2}}%
\tilde{F}(\tilde{\varphi})\right] ,  \label{32t}
\end{equation}%
which corresponds to an action for the $4$-dimensional gravity coupled
non-minimally to a scalar field. Note that this action has the form%
\begin{equation}
\tilde{S}_{B}=\tilde{S}_{g}+\tilde{S}_{g\varphi }+\tilde{S}_{\varphi },
\end{equation}%
where, $\tilde{S}_{g}$ is a pure gravitational action term, $\tilde{S}%
_{g\varphi }$ is a non-minimal interaction term between gravity and a scalar
field, and $\tilde{S}_{\varphi }$ represents a kind of scalar field
potential. In order to write down the action in the usual way, we define the
constant $\varepsilon $ and the potential $V(\varphi )$ as (removing the
symbols $\sim $ in (\ref{32t})). In fact 
\begin{equation}
\varepsilon =\frac{4\kappa r_{c}^{2}}{3}\text{ \ \ },\text{\ \ }V(\varphi )=%
\frac{3C}{4\kappa r_{c}^{2}}F(\varphi ),  \label{100t}
\end{equation}%
where $\kappa $ is the gravitational constant. This permits to rewrite the
action for a $4$-dimensional brane non-minimally coupled to a scalar field,
immersed in a $5$-dimensional space-time as%
\begin{equation}
S[g,\varphi ]=\int d^{4}x\sqrt{-g}\left[ R+\varepsilon RV(\varphi )-2\kappa
V(\varphi )\right] .  \label{33t}
\end{equation}%
The corresponding field equations describing the behavior of the $4$%
-dimensional brane in the presence of the scalar field $\varphi $ are given
by

\begin{equation}
G_{\mu \nu }\left( 1+\varepsilon V\right) +\varepsilon H_{\mu \nu }=-\kappa
g_{\mu \nu }V,  \label{z1}
\end{equation}%
\begin{equation}
\frac{\partial V}{\partial \varphi }\left( 1-\frac{\varepsilon R}{2\kappa }%
\right) =0,  \label{z2}
\end{equation}%
where%
\begin{equation}
H_{\mu \nu }=g_{\mu \nu }\nabla ^{\lambda }\nabla _{\lambda }V-\nabla _{\mu
}\nabla _{\nu }V.
\end{equation}

In order to construct a model of universe based on Eqs. (\ref{z1}-\ref{z2}),
we consider the Friedmann-Lema\^{\i}tre-Robertson-Walker metric%
\begin{equation}
ds^{2}=-dt^{2}+a^{2}(t)\left( \frac{dr^{2}}{1-kr^{2}}+r^{2}\left( d\theta
^{2}+\sin ^{2}\theta d\psi ^{2}\right) \right) ,
\end{equation}%
where $a(t)$ is the so called "cosmic scale factor" and $k=0,+1,-1$
describes flat, spherical and hyperbolic spatial geometries, respectively.
Following the usual procedure, we find the following field equations

\begin{eqnarray}
3\left( H^{2}+\frac{k}{a^{2}}\right) \left( 1+\varepsilon V\right)
+3\varepsilon H\dot{\varphi}\frac{\partial V}{\partial \varphi } &=&V,
\label{iii1} \\
\left( 2\dot{H}+3H^{2}+\frac{k}{a^{2}}\right) \left( 1+\varepsilon V\right)
+\varepsilon \left( \dot{\varphi}^{2}\frac{\partial ^{2}V}{\partial \varphi
^{2}}+\left( \ddot{\varphi}+2H\dot{\varphi}\right) \frac{\partial V}{%
\partial \varphi }\right) &=&V,  \label{iii2} \\
\frac{\partial V}{\partial \varphi }\left[ 1-3\varepsilon \left( \dot{H}%
+2H^{2}+\frac{k}{a^{2}}\right) \right] &=&0,  \label{iii3}
\end{eqnarray}%
where $H=\dot{a}/a$ is the Hubble parameter and we have used natural unities 
$\kappa =8\pi G=c=1$. Dot means derivative with respect to time.

From (\ref{iii1}, \ref{iii2}, \ref{iii3}) we see that when $\varepsilon =0$
and $V=$ $const.$, we have a de Sitter behavior governed by $H=\sqrt{V/3}.$
On the another hand, from equation (\ref{100t}) we can see that $\varepsilon 
$ is a positive quantity. This fact allows us to define an effective
cosmological constant as%
\begin{equation}
\Lambda _{eff}=\frac{1}{2\varepsilon },
\end{equation}%
which will play an important role in the cosmological consequences that we
will show below.

In the flat case, the Eqs. (\ref{iii1}, \ref{iii2}, \ref{iii3}) are given by

\begin{eqnarray}
3H^{2} &=&-\frac{1}{V+2\Lambda _{eff}}\left( 3H\frac{dV}{dt}-2\Lambda
_{eff}V\right) ,  \label{iii4} \\
2\dot{H}+3H^{2} &=&-\frac{1}{V+2\Lambda _{eff}}\left( \frac{d^{2}V}{dt^{2}}%
+2H\frac{dV}{dt}-2\Lambda _{eff}V\right) ,  \label{iii5}
\end{eqnarray}%
where the equation\textbf{\ (}\ref{iii3}) was not considered because it is
not an independent equation. In fact, subtracting\textbf{\ (}\ref{iii5})%
\textbf{\ }from\textbf{\ }(\ref{iii4})\textbf{\ }we obtain%
\begin{equation}
2\dot{H}=-\frac{1}{V+2\Lambda _{eff}}\left( \frac{d^{2}V}{dt^{2}}-H\frac{dV}{%
dt}\right) .  \label{iii5'}
\end{equation}%
Deriving\textbf{\ (}\ref{iii4}\textbf{) }with respect to time and using 
\textbf{(}\ref{iii5'})\textbf{\ }we find\textbf{\ (}\ref{iii3}), when\textbf{%
\ }$k=0$\textbf{. }Bear in mind that, at the end of this Section, we will
study an interesting consequence derived from this equation.

We write now the Eqs. (\ref{iii4}, \ref{iii5}) in the "standard" form%
\begin{eqnarray}
3H^{2} &=&\rho \text{ \ \ },\text{ \ }\rho =-\frac{1}{V+2\Lambda _{eff}}%
\left( 3H\frac{dV}{dt}-2\Lambda _{eff}V\right) ,  \label{iii6} \\
\dot{H}+H^{2} &=&-qH^{2}=-\frac{1}{6}\left( \rho +3p\right) ,  \notag \\
\frac{1}{6}\left( \rho +3p\right) &=&\frac{1}{4\Lambda _{eff}}\frac{1}{%
\left( 1+V/2\Lambda _{eff}\right) }\left( \frac{d^{2}V}{dt^{2}}+H\frac{dV}{dt%
}-\frac{4\Lambda _{eff}}{3}V\right) ,  \label{iii7}
\end{eqnarray}%
being $q$ the deceleration parameter defined by $q=-1-\dot{H}/H^{2}$ and $p$
the pressure associated to $\rho $ given by

\begin{equation}
p=\frac{1}{V+2\Lambda _{eff}}\left( \frac{d^{2}V}{dt^{2}}+2H\frac{dV}{dt}%
-2\Lambda _{eff}V\right) ,  \label{iii8}
\end{equation}%
which allows to write the barotropic equation $p=\omega \rho $, where

\begin{equation}
\omega =-\left( \frac{2\Lambda _{eff}V-2HdV/dt-d^{2}V/dt^{2}}{2\Lambda
_{eff}V-3HdV/dt}\right) ,  \label{iii9}
\end{equation}%
and we note that $V=const.$ leads to $\omega =-1$, i.e., a de Sitter
evolution.

Considering again Eqs. (\ref{iii4}, \ref{iii5}) and defining\ $x=V/2\Lambda
_{eff}$,\ we write the field equations in the form

\begin{eqnarray}
3\left( H^{2}+H\frac{d}{dt}\ln \left( 1+x\right) \right) &=&2\Lambda
_{eff}\left( \frac{x}{1+x}\right) ,  \label{iii11} \\
3\left( qH^{2}-H\frac{d}{dt}\ln \left( 1+x\right) \right) &=&-2\Lambda
_{eff}\left( \frac{x}{1+x}\right) +\frac{3}{2}\left( \frac{1}{1+x}\right) 
\frac{d^{2}x}{dt^{2}},  \label{iii12}
\end{eqnarray}%
and we discuss some examples:

\textbf{(a) }$x=x_{0}$\ $=$\textbf{\ }const.\textbf{\ }If $x$\ behaves as a
constant\textbf{,} the solution for the Hubble parameter is given by 
\begin{equation}
H=\sqrt{\frac{2\Lambda _{eff}}{3}\frac{x_{0}}{1+x_{0}}},  \label{iii14}
\end{equation}%
i.e., a de Sitter evolution for all time.

\textbf{(b)} $x=t/t_{0}$. In this case, the solution is\textit{\ } 
\begin{equation}
H\left( t\right) =\frac{1}{2t_{0}\left( 1+t/t_{0}\right) }\left( \sqrt{1+%
\frac{8\Lambda _{eff}t_{0}}{3}\left( 1+t/t_{0}\right) t}-1\right) ,
\label{iii15}
\end{equation}%
where we can see that 
\begin{equation}
H\left( t\rightarrow \infty \right) \rightarrow \sqrt{\frac{2\Lambda _{eff}}{%
3}}\ \text{\ \ },\text{\ \ }\rho \left( t\rightarrow \infty \right)
\rightarrow 2\Lambda _{eff}\text{ \ \ \ }and\text{\ \ \ }q\left(
t\rightarrow \infty \right) =-1,  \label{iii16}
\end{equation}%
which means that we have a late de Sitter evolution.

\textbf{(c) }$x=$ $\exp \left( t/t_{0}\right) $. Here, the Hubble parameter
turns out to be 
\begin{equation}
H\left( t\right) =\frac{1}{2t_{0}\left[ 1+\exp \left( t/t_{0}\right) \right] 
}\left( \sqrt{1+\frac{8\Lambda _{eff}t_{0}^{2}}{3}\exp \left( t/t_{0}\right) %
\left[ 1+\exp \left( t/t_{0}\right) \right] }-1\right) ,
\end{equation}%
and 
\begin{equation}
\text{ \ \ \ }H\left( t\rightarrow \infty \right) \rightarrow \sqrt{\frac{%
2\Lambda _{eff}}{3}}\text{ \ \ },\text{ \ }\rho \left( t\rightarrow \infty
\right) \rightarrow 2\Lambda _{eff}\text{ \ \ }and\text{ \ \ }q\left(
t\rightarrow \infty \right) \rightarrow -1,
\end{equation}%
and, as in the previous case, we have a late de Sitter evolution.

From (\ref{iii9}) the parameter of the equation of state takes the form

\begin{equation}
\omega \left( t\right) =-\left( \frac{2\Lambda _{eff}x-2Hdx/dt-d^{2}x/dt^{2}%
}{2\Lambda _{eff}x-3Hdx/dt}\right) ,  \label{iii19}
\end{equation}%
and if $x=t/t_{0}$ one finds%
\begin{equation}
\omega \left( t\right) =-\left( \frac{t-H\left( t\right) /\Lambda _{eff}}{%
t-3H\left( t\right) /2\Lambda _{eff}}\right) ,
\end{equation}%
so that if we identify $t_{0}$\ as the current time, then

\begin{equation}
\text{\ }\omega \left( t_{0}\right) =-\left( \frac{t_{0}-H\left(
t_{0}\right) /\Lambda _{eff}}{t_{0}-3H\left( t_{0}\right) /2\Lambda _{eff}}%
\right) <-1\text{ \ \ \ }and\text{ \ \ \ }\omega \left( t\rightarrow \infty
\right) =-1,\text{ \ \ \ }
\end{equation}%
and we have a transient phantom evolution (not ruled out by the current
observational data). Theoretical frameworks where this type of evolution is
discussed can be seen in \cite{lepe1}, \cite{zar}, and \cite{lepe2}.

The shown examples above have a common characteristic, namely they show a
late de Sitter evolution like, for instance, $\Lambda CDM$ at late times,
but we do not know if this characteristic comes from the formalism that we
are inspecting or from the choice (Ansatz) that we make for $V\left(
t\right) $. Since we do not have something to guide us towards a form for $%
V\left( t\right) $ from first principles, we are tied to playing with
different Ansatze for that potential. At least those shown here, give us
interesting results, in particular, that obtained from the Ansatz given in ($%
b$), a transient phantom evolution.

Previously, we have seen that equation\textbf{\ }(\ref{iii3})\textbf{\ }is
not independent, and therefore it was not analyzed in the first instance.
However, it reveals an interesting fact of our scheme, the presence of a
cosmological bounce. In case that\textbf{\ }$k=0$\textbf{,\ }$\Lambda
_{eff}= $ $1/2\varepsilon $ and $\partial V/\partial \varphi \neq 0$,\textbf{%
\ }the equation\textbf{\ }(\ref{iii3})\textbf{\ }takes the form

\begin{equation}
\frac{2}{3}\Lambda _{eff}-\left( \dot{H}+2H^{2}\right) =0,  \label{v1}
\end{equation}%
which leads to the following solution for the Hubble parameter

\begin{equation}
H\left( t\right) =\sqrt{\Lambda _{eff}/3}\left( \frac{\exp \left[ 4\sqrt{%
\Lambda _{eff}/3}\left( t-t_{0}\right) \right] -1/\Delta \left( t_{0}\right) 
}{\exp \left[ 4\sqrt{\Lambda _{eff}/3}\left( t-t_{0}\right) \right]
+1/\Delta \left( t_{0}\right) }\right) ,  \label{v2}
\end{equation}%
where%
\begin{equation}
\Delta \left( t_{0}\right) =\frac{\sqrt{\Lambda _{eff}/3}+H_{0}}{\sqrt{%
\Lambda _{eff}/3}-H_{0}}.
\end{equation}%
Note that the equation (\ref{v2})\textbf{\ }can be written in terms of an
hyperbolic tangent as%
\begin{equation}
H\left( t\right) =\sqrt{\frac{\Lambda _{eff}}{3}}\tanh \left( 2\sqrt{\frac{%
\Lambda _{eff}}{3}}\left( t-t_{0}\right) +\frac{1}{2}\ln \left[ \Delta
\left( t_{0}\right) \right] \right) ,  \label{v3}
\end{equation}%
which reveals a cosmological bounce in%
\begin{equation}
t_{b}=t_{0}-\frac{1}{4}\sqrt{\frac{3}{\Lambda _{eff}}}\ln \Delta \left(
t_{0}\right) \Longrightarrow H\left( t_{b}\right) =0.  \label{v4}
\end{equation}%
Moreover, the expression (\ref{v3}) is also showing that $H<0$ for $t<t_{b}$%
, and $H>0$ for $t>t_{b}$; i.e., there is a contraction for $t<t_{b}$ and an
expansion for $t>t_{b}$. In fact, from equation (\ref{v3}) the cosmic scale
factor is obtained%
\begin{equation}
a(t)=a_{b}\cosh ^{1/2}\left( 2\sqrt{\frac{\Lambda _{eff}}{3}}\left(
t-t_{b}\right) \right) ,  \label{v5}
\end{equation}%
where $a_{b}$ is the minimum of the scale factor, which occurs at $t_{b}$,
and it is given by%
\begin{equation}
a_{b}=a(t_{b})=a\left( 0\right) \cosh ^{-1/2}\left( 2\sqrt{\frac{\Lambda
_{eff}}{3}}\left( t_{0}-t_{b}\right) \right) .  \label{v6}
\end{equation}%
Behaviors of this kind have been frequently studied in the literature in the
context of cosmological bouncing; see e.g., \cite{Biswas1}, \cite{Biswas}, 
\cite{Bamba}. Finally, we note that 
\begin{equation}
H(t\rightarrow \infty )\rightarrow \sqrt{\frac{\Lambda _{eff}}{3}},
\label{v7}
\end{equation}%
i.e., a late de Sitter evolution.

\section{\textbf{Introduction of the kinetic term }$\left( 1/2\right) \dot{%
\protect\varphi}^{2}$}

In Ref. \cite{salg4} it was found that the surface term $B_{\text{EChS}%
}^{(4)}$ in the Lagrangian (\ref{1t}) is given by%
\begin{align}
B_{EChS}^{(4)}& =\alpha _{1}l^{2}\epsilon _{abcde}e^{a}\omega ^{bc}\left( 
\frac{2}{3}d\omega ^{de}+\frac{1}{2}\omega _{\text{ \ }f}^{d}\omega
^{fe}\right)  \notag \\
& \quad +\alpha _{3}\epsilon _{abcde}\left[ l^{2}\left( h^{a}\omega
^{bc}+k^{ab}e^{c}\right) \left( \frac{2}{3}d\omega ^{de}+\frac{1}{2}\omega _{%
\text{ \ }f}^{d}\omega ^{fe}\right) \right.  \notag \\
& \qquad \qquad \qquad \left. +l^{2}k^{ab}\omega ^{cd}\left( \frac{2}{3}%
de^{e}+\frac{1}{2}\omega _{\text{ \ }f}^{d}e^{e}\right) +\frac{1}{6}%
e^{a}e^{b}e^{c}\omega ^{de}\right] .  \label{8'}
\end{align}

From (\ref{1t}) and (\ref{8'}) we can see that kinetic terms corresponding
to the fields $h^{a}$ and $k^{ab}$, absent in the Lagrangian, are present in
the surface term. This situation is common to all Chern-Simons theories.
This has the consequence that the action (\ref{33t}) does not have the
kinetic term for the scalar field $\varphi $.

It could be interesting to add a kinetic term to the $4$-dimensional brane
action. In this case, the action (\ref{33t}) takes the form%
\begin{equation}
S[g,\varphi ]=\int d^{4}x\sqrt{-g}\left[ R+\varepsilon RV(\varphi )-2\kappa %
\left[ \frac{1}{2}\left( \nabla _{\mu }\varphi \right) \left( \nabla ^{\mu
}\varphi \right) +V(\varphi )\right] \right] .  \label{34t}
\end{equation}%
The corresponding field equations are given by

\begin{equation}
G_{\mu \nu }\left( 1+\varepsilon V\right) +\varepsilon H_{\mu \nu }=\kappa
T_{\mu \nu }^{\varphi },  \label{40t}
\end{equation}%
\begin{equation}
\nabla _{\mu }\nabla ^{\mu }\varphi -\frac{\partial V}{\partial \varphi }%
\left( 1-\frac{\varepsilon R}{2\kappa }\right) =0,  \label{41t}
\end{equation}%
where $T_{\mu \nu }^{\varphi }$ is the energy-momentum tensor of the scalar
field%
\begin{equation}
T_{\mu \nu }^{\varphi }=\nabla _{\mu }\varphi \nabla _{\nu }\varphi -g_{\mu
\nu }\left( \frac{1}{2}\nabla ^{\lambda }\varphi \nabla _{\lambda }\varphi
+V\right) ,  \label{37t}
\end{equation}%
and the rank-2 tensor $H_{\mu \nu }$ is defined as%
\begin{equation}
H_{\mu \nu }=g_{\mu \nu }\nabla ^{\lambda }\nabla _{\lambda }V-\nabla _{\mu
}\nabla _{\nu }V.  \label{38t}
\end{equation}

Following the usual procedure, we find that the FLRW type equations are
given by%
\begin{equation}
3\left( \frac{\dot{a}^{2}+k}{a^{2}}\right) \left( 1+\varepsilon V\right)
+3\varepsilon \frac{\dot{a}}{a}\dot{\varphi}\frac{\partial V}{\partial
\varphi }=\kappa \left( \frac{1}{2}\dot{\varphi}^{2}+V\right) ,  \label{48x}
\end{equation}%
\begin{equation}
\left( 2\frac{\ddot{a}}{a}+\frac{\dot{a}^{2}+k}{a^{2}}\right) \left(
1+\varepsilon V\right) +\varepsilon \left[ \dot{\varphi}^{2}\frac{\partial
^{2}V}{\partial \varphi ^{2}}+\left( \ddot{\varphi}+2\frac{\dot{a}}{a}\dot{%
\varphi}\right) \frac{\partial V}{\partial \varphi }\right] =-\kappa \left( 
\frac{1}{2}\dot{\varphi}^{2}-V\right) ,  \label{49x}
\end{equation}%
\begin{equation}
\ddot{\varphi}+3\frac{\dot{a}}{a}\dot{\varphi}+\frac{\partial V}{\partial
\varphi }\left[ 1-\frac{3\varepsilon }{\kappa }\left( \frac{\ddot{a}}{a}+%
\frac{\dot{a}^{2}+k}{a^{2}}\right) \right] =0.  \label{50x}
\end{equation}%
In the case $k=0$, \textbf{and using }$\kappa =1$, Eqs. (\ref{48x},\ref{49x},%
\ref{50x}) takes the form

\begin{eqnarray}
3H^{2}\left( 1+\epsilon V\right) +3\epsilon H\frac{dV}{dt} &=&\frac{1}{2}%
\dot{\varphi}^{2}+V,  \label{iv1} \\
\left( 2\dot{H}+3H^{2}\right) \left( 1+\epsilon V\right) +\epsilon \left( 
\frac{d^{2}V}{dt^{2}}+2H\frac{dV}{dt}\right) &=&-\left( \frac{1}{2}\dot{%
\varphi}^{2}-V\right) ,  \label{iv2} \\
\left( \ddot{\varphi}+3H\dot{\varphi}\right) \dot{\varphi}+\frac{dV}{dt}%
\left[ 1-3\epsilon \left( \dot{H}+2H^{2}\right) \right] &=&0,  \label{iv3}
\end{eqnarray}%
and here, the equation\textbf{\ }(\ref{iv3})\textbf{\ }it is not an
independent equation. In fact, subtracting the equation\textbf{\ }(\ref{iv2})%
\textbf{\ }from the\textbf{\ }equation\textbf{\ }(\ref{iv1})\textbf{\ }we
obtain%
\begin{equation}
2\dot{H}\left( 1+\varepsilon \right) =\varepsilon H\frac{dV}{dt}-\varepsilon 
\frac{d^{2}V}{dt^{2}}+\dot{\varphi}^{2}.  \label{iv3'}
\end{equation}%
Deriving the equation\textbf{\ }(\ref{iv1})\textbf{\ }with respect to time
and using the equation\textbf{\ }(\ref{iv3'})\textbf{\ }we find the equation%
\textbf{\ }(\ref{iv3})\textbf{.}

The combination $\left( 1/2\right) \dot{\varphi}^{2}+V\left( \varphi \right) 
$, together with the combination $\left( 1/2\right) \dot{\varphi}%
^{2}-V\left( \varphi \right) $, reminds us that in a standard scalar field
theory

\begin{equation}
\rho =\frac{1}{2}\dot{\varphi}^{2}+V\text{ \ },\text{ \ }p=\frac{1}{2}\dot{%
\varphi}^{2}-V,  \label{iv4}
\end{equation}%
which is recovered at the limit $\varepsilon \rightarrow 0$. In fact, when $%
\varepsilon \rightarrow 0$ the equations (\ref{iv1}, \ref{iv2}, \ref{iv3})%
\textbf{\ }takes the form%
\begin{eqnarray}
3H^{2} &=&\rho ,  \label{iv5} \\
\text{ \ }2\dot{H}+3H^{2} &=&-p, \\
\dot{\rho}+3H\left( p+\rho \right) &=&0,  \label{iv7}
\end{eqnarray}%
where we have used\textbf{\ }(\ref{iv4})\textbf{\ }to obtain\textbf{\ }(\ref%
{iv7})\textbf{\ }from\textbf{\ }(\ref{iv3})\textbf{\ }when\textbf{\ }$%
\varepsilon \rightarrow 0$\textbf{.}

The equation (\ref{iv4}) allows us to write (\ref{iv1},\ref{iv2}) in the form%
\begin{eqnarray}
3H^{2} &=&\frac{1}{1+\varepsilon V}\left( \rho -3\varepsilon H\dot{\varphi}%
\frac{\partial V}{\partial \varphi }\right) ,  \label{iv17} \\
2\dot{H}+3H^{2} &=&-\frac{1}{1+\varepsilon V}\left( p-\varepsilon \left[
\left( \rho +p\right) \frac{\partial ^{2}V}{\partial \varphi ^{2}}+\left( 
\ddot{\varphi}+2H\dot{\varphi}\right) \frac{\partial V}{\partial \varphi }%
\right] \right) ,  \label{iv18}
\end{eqnarray}%
and if we choose $p=-\rho $, by thinking in a de Sitter evolution, we obtain%
\begin{equation}
3H^{2}=\frac{\rho }{1+\varepsilon V}-3H\frac{d}{dt}\ln \left( 1+\varepsilon
V\right) ,  \label{(a)}
\end{equation}%
\begin{equation}
\dot{H}=\frac{1}{2}\left( \frac{\ddot{\varphi}}{\dot{\varphi}}+5H\right) 
\frac{d}{dt}\ln \left( 1+\varepsilon V\right) .  \label{(b)}
\end{equation}%
According to (\ref{(b)}), $\dot{H}=0$ say us

\begin{equation}
\frac{\ddot{\varphi}}{\dot{\varphi}}+5H=0\text{ \ \ }or\text{ \ \ }V\left(
t\right) =const.,
\end{equation}%
and, according to (\ref{(a)}), $V\left( t\right) =const.$ implies $\rho
=const.$ i.e., $H=const$., i.e., an usual de Sitter evolution. But, if $%
\ddot{\varphi}/\dot{\varphi}+5H=0$ and $V\left( t\right) \neq const.$, after
to see (\ref{(a)}) we have, with $H=H_{0}=const$.,

\begin{equation}
\rho \left( t\right) =3H_{0}\left[ H_{0}+\frac{d}{dt}\ln \left(
1+\varepsilon V\left( t\right) \right) \right] \left( 1+\varepsilon V\left(
t\right) \right) ,
\end{equation}%
and we have a de Sitter evolution although $\rho \neq const$. One more
detail, the equation $\ddot{\varphi}/\dot{\varphi}+5H=0$ has the solution $%
\dot{\varphi}\sim a^{-5}$ and so, the kinetic term $\left( 1/2\right) \dot{%
\varphi}^{2}$ dissolves very quickly with evolution leading us to $\rho \sim
V\left( t\right) $ and $p\sim -V\left( t\right) $ at late times, i.e., a de
Sitter evolution.

Writing (\ref{iv17},\ref{iv18}) in the form

\begin{eqnarray}
3H^{2} &=&\frac{1}{1+\varepsilon V}\left( \rho -3\varepsilon H\frac{dV}{dt}%
\right) ,  \label{iv25} \\
\dot{H}+H^{2} &=&-\frac{1}{6\left( 1+\varepsilon V\right) }\left[ \left(
\rho +3p\right) +3\varepsilon \left( \frac{d^{2}V}{dt^{2}}+H\frac{dV}{dt}%
\right) \right] ,  \label{iv26}
\end{eqnarray}%
we see that when $\varepsilon =0$ we recover the results of General
Relativity, i.e., $3H^{2}=\rho $ and $\dot{H}+H^{2}=-\left( 1/6\right)
\left( 1+3\omega \right) \rho $. By following this reminder, we write (\ref%
{iv25}, \ref{iv26}) in the standard form

\begin{eqnarray}
3H^{2} &=&\rho _{tot}, \\
\text{ \ \ }\dot{H}+H^{2} &=&-\frac{1}{6}\left( \rho _{tot}+3p_{tot}\right) ,
\\
\rho _{tot} &=&\frac{1}{1+\varepsilon V}\left( \rho -3H\frac{d\left(
\varepsilon V\right) }{dt}\right) , \\
\text{\ \ }p_{tot} &=&\frac{1}{1+\varepsilon V}\left( p+2H\frac{d\left(
\varepsilon V\right) }{dt}+\frac{d^{2}\left( \varepsilon V\right) }{dt^{2}}%
\right) ,  \label{iv28}
\end{eqnarray}%
we can see that we can build a barotropic equation $p_{tot}=\omega
_{tot}\rho _{tot}$, where

\begin{equation}
\omega _{tot}=\frac{p+2Hd\left( \varepsilon V\right) /dt+d^{2}\left(
\varepsilon V\right) /dt^{2}}{\rho -3Hd\left( \varepsilon V\right) /dt}\text{
\ \ }and\text{ \ \ }q=\frac{1}{2}\left( 1+3\omega _{tot}\right) .
\label{iv29}
\end{equation}%
By doing $p=\omega \rho $, we can write (\ref{iv29}) as

\begin{equation}
\omega _{tot}=\omega +\frac{\left( 2+3\omega \right) Hd\left( \varepsilon
V\right) /dt+d^{2}\left( \varepsilon V\right) /dt^{2}}{\rho -3Hd\left(
\varepsilon V\right) /dt}.  \label{iv30}
\end{equation}%
Here we can see that if $d^{2}\left( \varepsilon V\right) /dt^{2}=0$ and $%
d\left( \varepsilon V\right) /dt\neq 0$, then $\omega =-2/3$, $\omega
_{tot}=-2/3$ and $q=-1/2.$ This means that $\omega _{tot}$ belongs to the
quintessence zone. So, with $\left( \alpha ,\beta \right) $ constants, $%
\varepsilon V\left( t\right) =\alpha \left( t/t_{0}\right) +\beta $ is an
obvious choice for $\varepsilon V\left( t\right) $.

On the other hand, it is direct to show that%
\begin{equation}
\text{ \ }\dot{\rho}_{tot}+3H\left( 1+\omega _{tot}\right) \rho _{tot}=0,
\label{iv32}
\end{equation}%
so that%
\begin{eqnarray}
\rho _{tot}\left( a\right) &=&\rho _{tot}\left( a_{0}\right) \left( \frac{%
a_{0}}{a}\right) ^{3}\exp \left( -3\int_{t_{0}}^{t}\omega _{tot}\left(
t\right) d\ln a\right) ,  \label{iv33} \\
and\text{ \ \ \ }\omega _{tot} &=&-\frac{2}{3}\rightarrow \rho _{tot}\left(
a\right) =\rho _{tot}\left( a_{0}\right) \left( \frac{a_{0}}{a}\right) , 
\notag
\end{eqnarray}%
and the same is true for $\rho \left( a\right) $, that is, $\rho \left(
a\right) =\rho \left( 0\right) \left( a_{0}/a\right) $.

Note that, if $V\left( t\right) =V_{0}=const.$%
\begin{equation}
\rho _{tot}=\frac{1}{1+\epsilon V_{0}}\rho \text{ \ \ },\text{ \ \ }p_{tot}=%
\frac{1}{1+\epsilon V_{0}}p\text{ \ \ \ }and\text{ \ \ }\omega _{tot}=\omega
,
\end{equation}%
and if $\omega =0$ then $\omega _{tot}=0$ and then $p_{tot}=0$. This means
that $\omega _{tot}=0$ plays the role of the usual dark matter ($\omega =0$%
), although $\rho _{tot}\neq \rho $.

Finally, we have been using the quantity\textbf{\ }$\Lambda
_{eff}=1/2\varepsilon $\textbf{, }where\textbf{\ }$\varepsilon =4\kappa
r_{c}^{2}/3=const.$\textbf{\ }is a parameter derived from the mechanism of
dimensional reduction under consideration, which depends on the
gravitational constant\textbf{\ }$\kappa $\textbf{\ }and the
compactification radius\textbf{\ }$r_{c}$\textbf{. }This parameter plays the
role of an effective cosmological constant (its inverse) recalling that in
the action\textbf{\ }$S[g,\varphi ]=\int d^{4}x\sqrt{-g}\left[ R+\left(
\varepsilon R-2\kappa \right) V(\varphi )\right] $\textbf{\ }there is no a
"bare" cosmological constant. This fact could lead us to conjecture that the%
\textbf{\ }$h$\textbf{-}field\textbf{\ (}or\textbf{\ }$\tilde{h}$\textbf{-}%
field\textbf{), }in some way, manifests itself as dark energy. If so, the
next step will be to submit the present outline to the verdict of
observation.

\section{\textbf{Concluding remarks}}

We have considered a modification of the Poincar\'{e} symmetries known as
given $\mathfrak{B}$ Lie algebra also known as generalized Poincar\'{e}
algebra, whose generators satisfy the commutation relation shown in Eq. (7)
of Ref. \cite{gomez}. Besides the vierbein $e_{\mu }^{a}$ and the spin
connection $\omega _{\mu }^{ab}$, our scheme includes the fields\textbf{\ }$%
k_{\mu }^{ab}$ and $h_{\mu }^{a}$ whose dynamic is described by the field
equation obtained from the corresponding actions.

We have used the field equations for a $4$-dimensional brane embedded in the 
$5$-dimensional spacetime of \cite{bra1} to study their cosmological
consequences. The corresponding FLRW equations are found by means of the
usual procedure and cosmological solutions are shown and discussed. We
highlight two solutions, by choosing $\partial V/\partial t=const.$, a
transient phantom evolution (not ruled out by the current observational
data) is obtained and if $\partial V/\partial t\neq const.$ we obtain a
bouncing solution.

Since the kinetic terms corresponding to the fields $h^{a}$ and $k^{ab}$ are
present in the surface term (see (\ref{1t}) and (\ref{8'})) it was necessary
to introduce a kinetic term to the $4$-dimensional action. As a consequence
of this, in the corresponding cosmological framework we highlight a de
Sitter evolution even when the energy density involved is not constant.

Whatever it is, and since we do not have something to guide us towards a
form for $V\left( t\right) $ from first principles, we are tied to playing
with different Ansatze for that potential. At least in the cases that were
considered give us interesting results. But, we must insist, we are
completely dependent on the Ansatze for $V\left( t\right) $. If we are
thinking on cosmology, the results shown here suffer from this "slavery".
The hope, a common feeling, is that what is shown can be a contribution that
guides us towards a better understanding of the present formalism and its
chance of being a possible alternative to General Relativity. It is evident
that the observational information will be key when it comes to
discriminating between both models\textbf{. }To extract information that
leads us to\textbf{\ }$V\left( t\right) $\textbf{\ }in order to visualize if
the scalar field\textbf{\ }philosophy has a viable chance of being real when
it comes to doing cosmology is the challenge to face.

\section{\textbf{Appendix. Derivation of the} \textbf{action for a }$4$%
\textbf{-dimensional brane embedded in the }$5$\textbf{-dimensional spacetime%
}}

In this Appendix we briefly review the derivation of the action\textbf{\ (%
\ref{31t'}). }In order to find it, we will first consider the following $5$%
-dimensional Randall-Sundrum \cite{randall} \cite{randall1} type metric%
\begin{eqnarray}
ds^{2} &=&e^{2f(\phi )}\tilde{g}_{\mu \nu }(\tilde{x})d\tilde{x}^{\mu }d%
\tilde{x}^{\nu }+r_{c}^{2}d\phi ^{2},  \notag \\
&=&\eta _{ab}e^{a}e^{b},  \notag \\
&=&e^{2f(\phi )}\tilde{\eta}_{mn}\tilde{e}^{m}\tilde{e}^{n}+r_{c}^{2}d\phi
^{2},  \label{5t2}
\end{eqnarray}%
where $e^{2f(\phi )}$ is the so-called "warp factor", and $r_{c}$ is the
so-called "compactification radius" of the extra dimension, which is
associated with the coordinate $0\leqslant \phi <2\pi $. The symbol $\sim $
denotes $4$-dimensional quantities$.$ We will use the usual notation%
\begin{eqnarray}
x^{\alpha } &=&\left( \tilde{x}^{\mu },\phi \right) ;\text{ \ \ \ \ }\alpha
,\beta =0,...,4;\text{ \ \ \ \ }a,b=0,...,4;  \notag \\
\mu ,\nu &=&0,...,3;\text{ \ \ \ \ }m,n=0,...,3;  \notag \\
\eta _{ab} &=&diag(-1,1,1,1,1);\text{ \ \ \ \ }\tilde{\eta}%
_{mn}=diag(-1,1,1,1),  \label{6t2}
\end{eqnarray}%
which allows us to write the vielbein 
\begin{equation}
e^{m}(\phi ,\tilde{x})=e^{f(\phi )}\tilde{e}^{m}(\tilde{x})=e^{f(\phi )}%
\tilde{e}_{\text{ }\mu }^{m}(\tilde{x})d\tilde{x}^{\mu };\text{ \ \ }%
e^{4}(\phi )=r_{c}d\phi ,\text{\ }  \label{s22}
\end{equation}%
where\textbf{\ }$\tilde{e}^{m}$\textbf{\ }is the vierbein.

From the vanishing torsion condition%
\begin{equation}
T^{a}=de^{a}+\omega _{\text{ }b}^{a}e^{b}=0,  \label{2t}
\end{equation}%
we obtain the connections 
\begin{equation}
\omega _{\text{ }b\alpha }^{a}=-e_{\text{ }b}^{\beta }\left( \partial
_{\alpha }e_{\text{ }\beta }^{a}-\Gamma _{\text{ }\alpha \beta }^{\gamma }e_{%
\text{ }\gamma }^{a}\right) ,  \label{3t}
\end{equation}%
where $\Gamma _{\text{ }\alpha \beta }^{\gamma }$ is the Christoffel symbol.

From Eqs. (\ref{s22}) and (\ref{2t}) we find%
\begin{equation}
\omega _{\text{ }4}^{m}=\frac{e^{f}f^{\prime }}{r_{c}}\tilde{e}^{m},
\label{102t}
\end{equation}%
and the $4$-dimensional vanishing torsion condition 
\begin{equation}
\tilde{T}^{m}=\tilde{d}\tilde{e}^{m}+\tilde{\omega}_{\text{ }n}^{m}\tilde{e}%
^{n}=0,  \label{1030t}
\end{equation}%
where\textbf{\ \ }$f^{\prime }=\frac{\partial f}{\partial \phi }$\textbf{, }$%
\tilde{\omega}_{\text{ }n}^{m}=\omega _{\text{ }n}^{m}$\textbf{\ }and\textbf{%
\ }$\tilde{d}=d\tilde{x}^{\mu }\frac{\partial }{\partial \tilde{x}^{\mu }}.$

From (\ref{102t}), (\ref{1030t}) and the Cartan's second structural
equation, $R^{ab}=d\omega ^{ab}+\omega _{\text{ }c}^{a}\omega ^{cb}$, we
obtain the components of the $2$-form curvature%
\begin{equation}
R^{m4}=\frac{e^{f}}{r_{c}}\left( f^{\prime 2}+f^{\prime \prime }\right)
d\phi \tilde{e}^{m},\text{ \ }R^{mn}=\tilde{R}^{mn}-\left( \frac{%
e^{f}f^{\prime }}{r_{c}}\right) ^{2}\tilde{e}^{m}\tilde{e}^{n},\text{\ }
\label{105t}
\end{equation}%
where the $4$-dimensional $2$-form curvature is given by%
\begin{equation}
\tilde{R}^{mn}=\tilde{d}\tilde{\omega}^{mn}+\tilde{\omega}_{\text{ }p}^{m}%
\tilde{\omega}^{pn}.
\end{equation}

The torsion-free condition implies that the third term in the EChS action,
given in equation (\ref{1t}), vanishes. This means that the corresponding
Lagrangian is no longer dependent on the field $k^{ab}$. So, the Lagrangian (%
\ref{1t}) has now two independent fields, $e^{a}$ and $h^{a}$, and it is
given by 
\begin{equation}
L_{ChS}^{(5)}[e,h]=\alpha _{1}l^{2}\varepsilon
_{abcde}R^{ab}R^{cd}e^{e}+\alpha _{3}\varepsilon _{abcde}\left( \frac{2}{3}%
R^{ab}e^{c}e^{d}e^{e}+l^{2}R^{ab}R^{cd}h^{e}\right) .  \label{4t}
\end{equation}

From Eq. (\ref{4t}) we can see that the Lagrangian contains the Gauss-Bonnet
term $L_{GB}$, the Einstein-Hilbert term $L_{EH}$ and a term $L_{H}$ which
couples geometry and matter. In fact, replacing (\ref{s22}) and (\ref{105t})
in (\ref{4t}) and using $\tilde{\varepsilon}_{mnpq}=\varepsilon _{mnpq4}$,
we obtain

\begin{eqnarray}
\tilde{S}[\tilde{e},\tilde{h}] &=&\int_{\Sigma _{4}}\tilde{\varepsilon}%
_{mnpq}\left( A\tilde{R}^{mn}\tilde{e}^{p}\tilde{e}^{q}+B\ \tilde{e}^{m}%
\tilde{e}^{n}\tilde{e}^{p}\tilde{e}^{q}+\right.  \notag \\
&&\left. +C\tilde{R}^{mn}\tilde{e}^{p}\tilde{h}^{q}+E\tilde{e}^{m}\tilde{e}%
^{n}\tilde{e}^{p}\tilde{h}^{q}\right) ,  \label{999}
\end{eqnarray}%
where%
\begin{equation}
h^{m}(\phi ,\tilde{x})=e^{g(\phi )}\tilde{h}^{m}(\tilde{x}),\text{ \ }%
h^{4}=0,  \label{o}
\end{equation}%
and\textbf{\ }%
\begin{equation}
A=2r_{c}\int_{0}^{2\pi }d\phi e^{2f}\left[ \alpha _{3}-\frac{\alpha _{1}l^{2}%
}{r_{c}^{2}}\left( 3f^{\prime 2}+2f^{\prime \prime }\right) \right] ,
\label{12t}
\end{equation}%
\begin{equation}
B=-\frac{1}{r_{c}}\int_{0}^{2\pi }d\phi e^{4f}\left[ \frac{2\alpha _{3}}{3}%
\left( 5f^{\prime 2}+2f^{\prime \prime }\right) -\frac{\alpha _{1}l^{2}}{%
r_{c}^{2}}f^{\prime 2}\left( 5f^{\prime 2}+4f^{\prime \prime }\right) \right]
,  \label{13t}
\end{equation}%
\begin{equation}
C=-\frac{4\alpha _{3}l^{2}}{r_{c}}\int_{0}^{2\pi }d\phi e^{f}e^{g}\left(
f^{\prime 2}+f^{\prime \prime }\right) ,  \label{14t}
\end{equation}%
\begin{equation}
E=\frac{4\alpha _{3}l^{2}}{r_{c}^{3}}\int_{0}^{2\pi }d\phi
e^{3f}e^{g}f^{\prime 2}\left( f^{\prime 2}+f^{\prime \prime }\right) ,
\label{15t}
\end{equation}%
with\textbf{\ }$f(\phi )$\textbf{\ }and\textbf{\ }$g(\phi )$\textbf{\ }%
representing functions that can be chosen (non-unique choice) as\textbf{\ }$%
f(\phi )=g(\phi )=ln(K+sin\phi )$\textbf{\ }with\textbf{\ }$K=constant>1$%
\textbf{; }and therefore we have%
\begin{equation}
A=\frac{2\pi }{r_{c}}\left[ \alpha _{3}r_{c}^{2}\left( 2K^{2}+1\right)
+\alpha _{1}l^{2}\right] ,  \label{201t}
\end{equation}%
\begin{equation}
B=\frac{\pi }{2r_{c}}\left[ \alpha _{3}\left( 4K^{2}+1\right) -\frac{\alpha
_{1}l^{2}}{2r_{c}^{2}}\right] ,  \label{21t}
\end{equation}%
\begin{equation}
C=-4r_{c}^{2}E=\frac{4\pi \alpha _{3}l^{2}}{r_{c}}.  \label{75t}
\end{equation}%
$\ \ \ \ \ \ \ $Taking into account that $L_{ChS}^{(5)}[e,h]$ flows into $%
L_{EH}^{(5)}$ when $l\longrightarrow 0$ \cite{irs}, we have that action (\ref%
{999}) should lead to the action of Einstein-Hilbert when $l\longrightarrow
0 $. From (\ref{999}) it is direct to see that this occurs when $A=-1/2$ and 
$B=0$. In this case, from Eqs. (\ref{201t}), (\ref{21t}) and (\ref{75t}), we
can see that%
\begin{equation}
\alpha _{1}=-\frac{r_{c}\left( 4K^{2}+1\right) }{2\pi l^{2}\left(
10K^{2}+3\right) },  \label{24t}
\end{equation}%
\begin{equation}
\alpha _{3}=-\frac{1}{4\pi r_{c}\left( 10K^{2}+3\right) },  \label{25t}
\end{equation}

\begin{equation}
C=-4r_{c}^{2}E=-\frac{l^{2}}{r_{c}^{2}\left( 10K^{2}+3\right) },
\end{equation}%
and therefore the action (\ref{999}) takes the form 
\begin{eqnarray}
\tilde{S}[\tilde{e},\tilde{h}] &=&\int_{\Sigma _{4}}\tilde{\varepsilon}%
_{mnpq}\left( -\frac{1}{2}\tilde{R}^{mn}\tilde{e}^{p}\tilde{e}^{q}\right. + 
\notag \\
&&\left. +C\tilde{R}^{mn}\tilde{e}^{p}\tilde{h}^{q}-\frac{C}{4r_{c}^{2}}%
\tilde{e}^{m}\tilde{e}^{n}\tilde{e}^{p}\tilde{h}^{q}\right) ,  \label{a}
\end{eqnarray}%
corresponding to a $4$-dimensional brane embedded in the $5$-dimensional
spacetime of the EChS gravity. We can see that when $l\rightarrow 0$ then $%
C\rightarrow 0$ and hence (\ref{a}) becomes the $4$-dimensional
Einstein-Hilbert action.

Finally, it is convenient to express the action\textbf{\ }(\ref{a})\textbf{\ 
}in tensorial language. To achieve this, we write\textbf{\ }$\tilde{e}^{m}(%
\tilde{x})=\tilde{e}_{\text{ }\mu }^{m}(\tilde{x})d\tilde{x}^{\mu }$\textbf{%
\ }and\textbf{\ }$\tilde{h}^{m}=\tilde{h}_{\text{ \ }\mu }^{m}d\tilde{x}%
^{\mu }$\textbf{, }and then we compute the individual terms in\textbf{\ }(%
\ref{a})\textbf{\ }as%
\begin{eqnarray}
\tilde{\varepsilon}_{mnpq}\tilde{R}^{mn}\tilde{e}^{p}\tilde{e}^{q} &=&-2%
\sqrt{-\tilde{g}}\tilde{R}d^{4}\tilde{x}, \\
\tilde{\varepsilon}_{mnpq}\tilde{R}^{mn}\tilde{e}^{p}\tilde{h}^{q} &=&2\sqrt{%
-\tilde{g}}\left( \tilde{R}\tilde{h}-2\tilde{R}_{\text{ }\nu }^{\mu }\tilde{h%
}_{\text{ }\mu }^{\nu }\right) d^{4}\tilde{x}, \\
\tilde{\varepsilon}_{mnpq}\tilde{e}^{m}\tilde{e}^{n}\tilde{e}^{p}\tilde{h}%
^{q} &=&6\sqrt{-\tilde{g}}\tilde{h}d^{4}\tilde{x},
\end{eqnarray}%
where it has been defined\textbf{\ }$\tilde{h}\equiv \tilde{h}_{\text{ }\mu
}^{\mu }$\textbf{. }So, the $4$-dimensional action for the brane immersed in
the $5$-dimensional space-time of the EChS gravitational theory is given by

\begin{equation}
\tilde{S}[\tilde{g},\tilde{h}]=\int d^{4}\tilde{x}\sqrt{-\tilde{g}}\left[ 
\tilde{R}+2C\left( \tilde{R}\tilde{h}-2\tilde{R}_{\text{ }\nu }^{\mu }\tilde{%
h}_{\text{ }\mu }^{\nu }\right) -\frac{3C}{2r_{c}^{2}}\tilde{h}\right] .
\end{equation}

\textbf{Acknowledgements}

This work was supported in part by\textit{\ }FONDECYT Grant\textit{\ }No.%
\textit{\ }1180681 from the Government of Chile. One of the authors (FG) was
supported by Grant \# R12/18 from Direcci\'{o}n de Investigaci\'{o}n,
Universidad de Los Lagos.

\end{document}